%% file: 0_main.tex
\documentclass{article}
\usepackage{spconf,amsmath,graphicx}
\usepackage{boldline}
\usepackage{array, makecell}
\usepackage{multicol}
\usepackage{multirow}
\usepackage{color}
\usepackage{booktabs}

\newcommand{\PreserveBackslash}[1]{\let\temp=\\#1\let\\=\temp}
\newcolumntype{C}[1]{>{\PreserveBackslash\centering}p{#1}}
\newcolumntype{R}[1]{>{\PreserveBackslash\raggedleft}p{#1}}
\newcolumntype{L}[1]{>{\PreserveBackslash\raggedright}p{#1}}

\newcommand{\figref}[1]{Fig.~\ref{#1}}
\newcommand{\tabref}[1]{Table.~\ref{#1}}

\newcommand{\equref}[1]{Eq.~(\ref{#1})}

\title{Phase continuity: Learning derivatives of phase spectrum\\ for Speech enhancement}

\name{Doyeon Kim$^{1}$\quad Hyewon Han$^{1}$\quad  Hyeon-Kyeong Shin$^{1,2}$\quad Soo-Whan Chung$^{2}$\quad Hong-Goo Kang$^{1}$}
\address{
{$^{1}$Dept. Electrical and Electronic Engineering, Yonsei University, South Korea}\newline \\
{$^{2}$Naver Corporation, South Korea}
}

\begin{document}

\maketitle

\input{1_abstract.tex}

\begin{keywords}
speech enhancement, denoising, phase reconstruction, phase continuity loss
\end{keywords}

\input{2_introduction.tex}
\input{3_related.tex}
\input{4_proposed.tex}
\input{5_experiments.tex}
\input{6_conclusion.tex}
\input{7_acknowledge.tex}
%
\

\vfill\pagebreak

\bibliographystyle{IEEEbib}
\bibliography{shortstrings,refs}

\end{document}

%% file: 1_abstract.tex
\begin{abstract}
Modern neural speech enhancement models usually include various forms of phase information in their training loss terms, either explicitly or implicitly.
However, these loss terms are typically designed to reduce the distortion of phase spectrum values at specific frequencies, which ensures they do not significantly affect the quality of the enhanced speech.
In this paper, we propose an effective phase reconstruction strategy for neural speech enhancement that can operate in noisy environments.
Specifically, we introduce a phase continuity loss that considers relative phase variations across the time and frequency axes.
By including this phase continuity loss in a state-of-the-art neural speech enhancement system trained with reconstruction loss and a number of magnitude spectral losses, we show that our proposed method further improves the quality of enhanced speech signals over the baseline, especially when training is done jointly with a magnitude spectrum loss.
\end{abstract}

%% file: 2_introduction.tex
\section{Introduction}
\label{sec:intro}
In many voice communication and voice-controlled interface systems, target speech signals are often distorted by background noise, resulting in uncomfortable communication and loss of intelligibility.
Many works have solved this problem by developing de-noising methods based on rigorous signal processing and deep learning techniques.

Conventionally, speech enhancement has been performed in the magnitude spectrum domain because of the assumption phase components are difficult to estimate in noisy environments.
However, enhancing only the magnitude spectrum has the limitation of reusing the distorted phase for speech reconstruction.
Recently, attention has been focused on incorporating spectral phase components to improve the performance of various speech-related tasks, such as improving speech intelligibility~\cite{paliwal2003usefulness} and speech recognition~\cite{schluter2001using,lindgren2003speech}.
Estimating phase information is challenging because there are no explicit ways to model the statistics of phase distortions caused by environmental factors.
Recent deep learning based studies have attempted to estimate phase information indirectly by minimizing the distance between the target and the estimated signals in the complex spectrum or waveform domain~\cite{erdogan2015phase,stoller2018wave}. 
However, these methods tend to put more weight on magnitude estimation rather than considering the phase information, which limits the role of the phase terms in the waveform reconstruction process.
To improve reconstruction performance, several studies~\cite{lee2019joint,zhang2021weighted} introduced additional phase loss terms during training to minimize the differences of wrapped phases between the target and output signals or construct another phase estimation network~\cite{yin2020phasen} to estimate the phase spectrum.

In this paper, we propose a novel training strategy for speech enhancement that considers the trajectory of phase components on both the time and frequency axes.
Unlike most conventional phase estimation methods that focus on instantaneous phase differences at each frequency bin, our proposed phase continuity loss considers phase variations in neighboring time frames and frequency bins.
To the best of our knowledge, this is the first method that uses a training criterion that utilizes both the time and frequency trajectories of the phase spectrum. 
The rationale behind our idea is as follows.
To improve the perceptual quality of enhanced speech signals, it is important to reconstruct voiced segments reliably, where phase continuity is a critical characteristic~\cite{mowlaee2014interspeech}.
Because the fundamental frequency and its harmonics may vary in consecutive analysis frames due to the dynamic nature of voicing, we consider phase differences between nearby time frames and frequency bins.
Considering the trade-off relationship between time and frequency resolutions, we also apply the proposed phase continuity loss to a number of phase spectra obtained by multi-spectral analysis techniques~\cite{yamamoto2020parallel}.
To verify the effectiveness of our strategy, we add our phase continuity loss into the training of a state-of-the-art speech enhancement network~\cite{defossez2020real} that used magnitude spectrum loss. Experimental results show that this improves the enhancement performance of the model when jointly used along with magnitude spectrum loss.

The remainder of the paper is organized as follows. Section~\ref{sec:related} describes several neural speech enhancement methods that utilize phase information.
We explain the learning strategy for the proposed method in Section~\ref{sec:proposed}, demonstrate the effectiveness in Section~\ref{sec:exp}, and draw conclusions in Section~\ref{sec:conclusion}.

%% file: 3_related.tex
\section{Related Works}
\label{sec:related}
Phase reconstruction in speech enhancement is a challenging but important task to improve perceptual quality and intelligibility~\cite{paliwal2003usefulness,alsteris2006further,paliwal2011importance}.
Recent deep learning techniques have accelerated phase-aware speech enhancement approaches by targeting the task of phase value estimation.
The phase-sensitive mask~(PSM)~\cite{erdogan2015phase} and complex ideal ratio mask~(cIRM)~\cite{williamson2015complex,hu2020dccrn} are two such examples.
However, because these methods reconstruct phase terms with the magnitude spectrum, they have the potential problem of learning phase information with imperfect context information.
To include phase distortion explicitly in the training objectives, \cite{lee2019joint} defines a phase loss using sinusoidal functions as follows:
\begin{equation}\label{eq:phaseloss}
    \mathcal{L}_{p}=\|\cos{\theta_{\mathrm{diff}}} - \cos{\hat{\theta}_{\mathrm{diff}}}\|_2 +\| \sin{\theta_{\mathrm{diff}}} - \sin{\hat{\theta}_{\mathrm{diff}}}\|_2,
\end{equation}
where $\theta_{\mathrm{diff}}$ and $\hat{\theta}_{\mathrm{diff}}$ are the unwrapped phase differences between the target and noisy input and between the target and enhanced speech, respectively.
Instead of estimating the phase components in the time-frequency domain, some methods directly enhance speech signal waveforms in the time domain using end-to-end training criteria. 
The basic objective is to reduce the difference between two waveforms using L1 or L2 distances~\cite{stoller2018wave}, signal-to-distortion ratio~(SDR) loss~\cite{venkataramani2017adaptive}, or scale-invariant signal-to-noise ratio (SI-SNR) loss~\cite{sun2021funnel} to maximize the cosine similarity between speech segments.
\cite{wang2015deep} introduces a frequency analysis approach to the training criteria, the short-time Fourier transform (STFT) loss, considering the spectrum consistency characteristics of speech.
Multi-resolution STFT loss~(MR-STFT)~\cite{yamamoto2020parallel} expands the STFT loss to multi-resolutions with different analysis frames, thereby implicitly providing phase information.
Recently, some GAN-based methods have proposed loss functions that guide the network to learn phase information by considering speech in multi-resolutions~\cite{yamamoto2020parallel,su2020hifi}.
A GAN-based method used objective evaluation metrics such as short-time objective intelligibility (STOI) and perceptual evaluation of speech quality (PESQ) scores for the discrimination task~\cite{fu2019metricgan}.
These GAN-based methods provide implicit guidance to learn many different characteristics of speech waveforms, including their phase components.

%% file: 4_proposed.tex
\begin{figure*}[t]
  \centering
  \centerline{\includegraphics[width=0.8\linewidth]{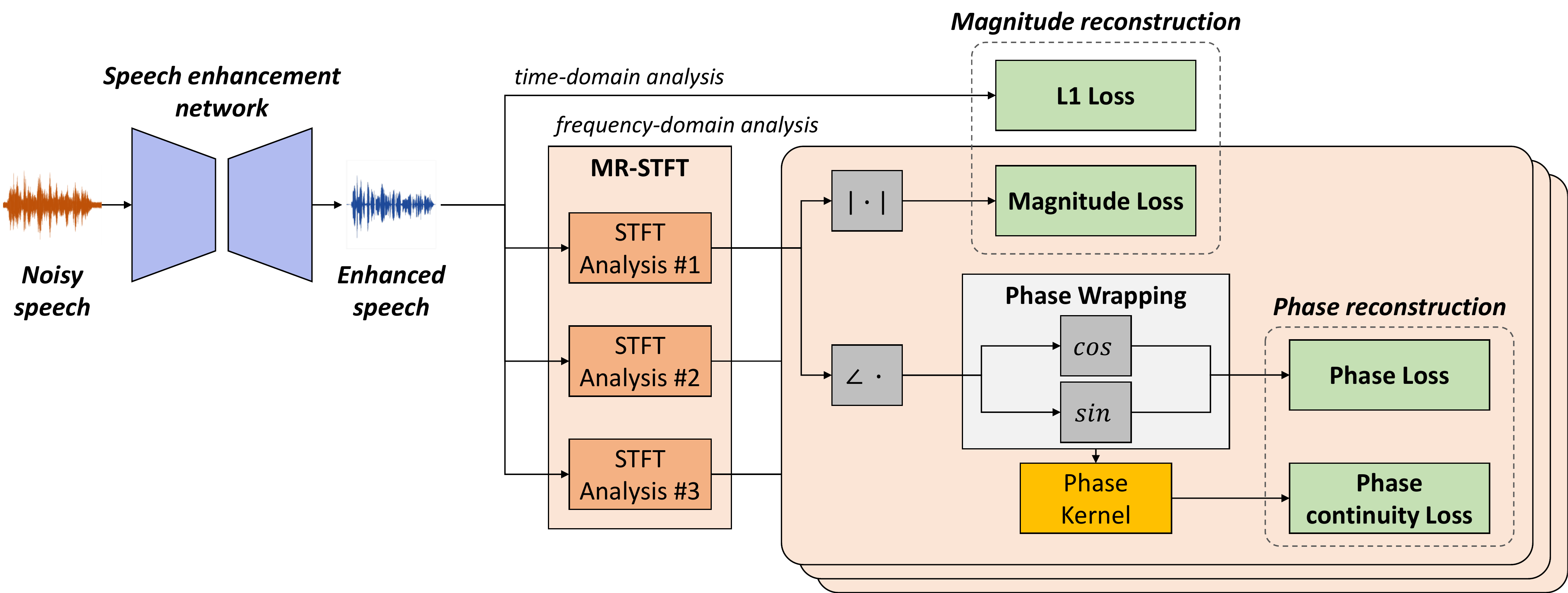}}
\vspace{-10 pt}
\caption{Illustration of the proposed phase-aware speech enhancement strategy.}
\label{fig:strategy}
\vspace{-15 pt}
\end{figure*}

\section{Proposed method}
\label{sec:proposed}

\subsection{Phase Continuity Loss}

The characteristics of a phase spectrum vary rapidly depending on local behaviors across the time and frequency axes.
In addition to the phase value itself, many works have shown the importance of phase derivatives across the time and frequency axes~\cite{mowlaee2014interspeech, mowlaee2016advances}.

In this work, we introduce a training criterion that considers both instantaneous frequency (IF) and group delay (GD) representations, which are phase derivatives along the time and frequency axes, respectively. We call this \textit{phase continuity loss (PCL)}.
The phase continuity physically indicates phase differences between neighboring values, 
and it implicitly presents enriched acoustic information such as pitch variation and the shape of the spectral envelope.
Our proposed training criterion minimizes the difference in the phase continuity between clean and enhanced speech.
If the resolution of the spectrum is infinitely high, the derivative of each phase component is defined as follows:
\vspace{-3 pt}
\begin{equation} \label{eq:diff}
    f'(\theta^k_{n}) = \lim_{i \rightarrow 0} \lim_{j\rightarrow 0}  \frac{f(\theta^k_{n})-f(\theta^{k+i}_{n+j})}{{\theta^k_n - \theta^{k+i}_{n+j}}},
\vspace{-3 pt}
\end{equation}
where $n$ and $k$ denote indices of the time and frequency bins, respectively.
We replace the phase terms using cosine and sine functions as a wrapping function like~\cite{lee2019joint}. The wrapping function in \equref{eq:diff} is $f(\theta)=cos(\theta)+sin(\theta)$.
\equref{eq:diff} represents IF and GD concurrently.
When $i$ and $j$ are small, it is important to consider the phase variations between neighboring time-frequency bins, including diagonally neighboring bins, to estimate the phase trajectories reliably (\textit{i.e.,} for the $(i,j)$ bin, we also consider the $(i\pm1,j\pm1)$ bins).

To make the processing easier, we construct a kernel to represent all of the derivatives, computing the differences between neighboring phase components.
This kernel captures the phase variations in the time and frequency axes simultaneously and can clearly represent the phase variation characteristics.
To calculate the broad range of phase continuity information, we build an $N\times N$ kernel, and set $N = 3$ in our experiments.
The detailed equation of the phase continuity kernel $\varphi(\cdot)$ is given as follows:
\vspace{-3 pt}
\begin{equation} \label{eq:one_kernel}
\begin{aligned}
    \varphi(\theta^n_k) =  \begin{bmatrix} f(\theta^{n-1}_{k+1}) & f(\theta^{n}_{k+1}) & f(\theta^{n+1}_{k+1}) \\ f(\theta^{n-1}_{k}) & f(\theta^{n}_{k}) & f(\theta^{n+1}_{k}) \\ f(\theta^{n-1}_{k-1}) & f(\theta^{n}_{k-1}) & f(\theta^{n+1}_{k-1})
\end{bmatrix}   - \textbf{1}  f(\theta^n_k),
\end{aligned}
\vspace{-3 pt}
\end{equation}
where $n$ and $k$ are the indices of time and frequency bins, respectively.
We use cosine and sine functions for $f(\cdot)$ to obtain wrapping results consistently with the phase loss term, which generates two continuity kernels $\varphi_{cos}(\theta)$ and $\varphi_{sin}(\theta)$.
The objective of PCL is to minimize the distance between the continuity kernels of the target and enhanced phase spectra:
\vspace{-3 pt}
\begin{equation}
    \mathcal{L}_{pc} = \|\varphi_{cos}(\theta) - \varphi_{cos}(\hat{\theta})\|_2 + \|\varphi_{sin}(\theta) - \varphi_{sin}(\hat{\theta})\|_2,
\vspace{-3 pt}
\end{equation}
where $\theta$ and $\hat{\theta}$ denote the phase value of the target and enhanced speech signals, respectively.

\subsection{Phase-aware training strategy}
Although the PCL is effective for learning phase variations and related information,
it is still challenging to reconstruct phase terms with only PCL due to training vulnerability.
Therefore, we train the model using both phase loss~(PL) and PCL terms.
Our overall phase-aware training criterion is as follows:
\vspace{-3 pt}
\begin{equation}
    \label{eq:loss}
    \mathcal{L}_{P} = \lambda_p \mathcal{L}_{p} + \lambda_{pc} \mathcal{L}_{pc},
\vspace{-3 pt}
\end{equation}
where $\mathcal{L}_{p}$ is similar to \equref{eq:phaseloss}, but with the wrapped phase before measuring the phase differences, and $\lambda_p$ and $\lambda_{pc}$ are weighting factors.
To consider various spectral characteristics, we obtain phase spectra with MR-STFT techniques.

Finally, we combine our proposed MR-PCL with a state-of-the-art speech enhancement network that originally only used multi-resolution magnitude spectrum loss.
\figref{fig:strategy} illustrates a block diagram of the overall proposed system and its training strategy.
The total training loss is as follows:
\vspace{-3 pt}
\begin{equation}
    \label{eq:totalloss}
    \mathcal{L}= \lambda_{0}\mathcal{L}_{L1} + \lambda_{1}\mathcal{L}_{STFT} + \lambda_{2}\mathcal{L}_{P},
\vspace{-3 pt}
\end{equation}
where $\mathcal{L}_{L1}$ denotes the loss using the L1 norm on waveform domain, while $\mathcal{L}_{STFT}$ and $\mathcal{L}_{P}$ are on the time-frequency domain with multi-resolution analysis.

%% file: 5_experiments.tex
\input{tables/table_snr}

\section{Experiments}
\label{sec:exp}
\subsection{Experimental settings}
\vspace{1pt}
\noindent\textbf{Data preparation.}
To train and evaluate the baseline and proposed methods, we used the VoiceBank-DEMAND~\cite{valentini2017noisy} dataset, which contains 16 kHz speech signals spoken by 30 speakers contaminated with 10 noise types.
This dataset comprises four types of SNRs for training and testing,~[0, 5, 10, 15] dB and [2.5, 7.5, 12.5, 17.5] dB each.
For data augmentation, we conducted pre-processing similar to the method described in \cite{defossez2020real} for a fair comparison. This step includes remixing noise, shifting audio samples, and masking band-stop filters.
To investigate the performance in harsh condition, we also evaluated the model on a dataset generated by remixing the noise and speech at $-5$ dB SNR.

\vspace{1pt}
\noindent\textbf{Network and training settings.}
We evaluated our training strategy on the DEMUCS model~\cite{defossez2020real}, an end-to-end speech enhancement model using the U-Net architecture with a long short-term memory (LSTM) network.
We re-implemented DEMUCS with 48 hidden layers, a stride size of 4, and an upsampling rate of 4.
For the multi-resolution analysis, we used the same settings as ~\cite{defossez2020real}.
Because estimating the spectral power term is more important for speech enhancement, we placed greater weight on the MR-STFT loss than the phase-related criteria.
When training the DEMUCS model with PL ($\lambda_2 = \lambda_p$), we set the loss weights as follows; $\lambda_0 : \lambda_1 : \lambda_2 = 0.02 : 1 : 1$.
When training with both PL and PCL, because the phase derivatives are more sensitive than the phase values, we set the weighting factor of PCL to be smaller than that of PL, using the weights  $\lambda_0 : \lambda_1 : \lambda_2 = 0.01 : 1 : 0.1$ with $\lambda_p : \lambda_{pc} = 1 : 0.5$.

\vspace{1pt}
\noindent\textbf{Evaluation metrics.}
We evaluated the speech enhancement performance of the baseline and the proposed methods using multiple objective measurements, including wideband PESQ (WB-PSEQ), STOI, extended STOI~(ESTOI), SDR improvement (SDRi), CSIG, CBAK, COVL, segmental SNR (SNRseg), frequency-weighted SNRseg (fwSNRseg), and normalized covariance measure (NCM).
WB-PESQ indicates the perceptual quality of speech signals in a wideband resolution, and CSIG, CBAK, and COVL are composite metrics reflecting mean-opinion-scores (MOS)~\cite{hu2007evaluation}.
Speech intelligibility is measured by STOI, ESTOI and NCM scores, which are related to linguistic expression.
SDRi, SNRseg, and fwSNRseg are related to signal reconstruction performance.

Furthermore, the effectiveness of the phase reconstruction is verified using phase-related metrics: unwrapped RMSE~(UnRMSE)~\cite{gaich2015speech1, mowlaee2016advances}, which measures phase-aware speech intelligibility, and phase derivative features-related values such as GD, and IF~\cite{gaich2015speech2}.
These metrics are measured by calculating the RMSE value between the enhanced speech and the ground-truth sample on the voiced frames where perceptually important compared to the non-speech region.

\subsection{Experimental results}
\noindent\textbf{Evaluation results using objective measurements.}
\tabref{table:denoising} summarizes our experimental results using the objective evaluation criteria.
For all metrics, the models trained with our multi-resolution phase-aware training strategy outperformed the baseline trained with only L1 and MR-STFT loss.
We see further improvement over all aspects of the baseline performance when phase-related loss terms are also included.
Specifically, learning phase derivatives increased overall scores more than learning phase spectrum directly.
In addition, we observed that the PCL helps stabilize the training process by reducing large fluctuations in errors caused by including a criterion in the direct phase value estimation. 
Notably, the WB-PESQ and SDRi scores of the proposed method significantly improved than the baseline by 0.067 and 0.187 points, respectively.
The PCL also showed improvement in all the composite measurements for MOS prediction.
Lastly, the incrementally higher SNRseg and fwSNRseg scores showed the effectiveness of our proposed method over the whole frequency bands on the denoising task.

\tabref{table:denoising_snr} presents the WB-PESQ, STOI, and SDRi results given in \tabref{table:denoising} in greater detail, giving the scores based on the various SNRs of the input signals.
Our phase-related training criteria improved perceptual quality and speech intelligibility at all SNR levels.
In particular, it provided a more effective approach to reconstruct the phase information of the speech signals than the conventional phase loss that directly estimates the phase spectrum at each frequency.
It is also interesting that PCL improves the scores even in low SNR cases, which is regarded as a challenging task.

Moreover, we evaluated our method in a harsh condition without further training in SNRs $<$ 0dB.
The results are given in~\tabref{table:5dB_result}, which shows a similar tendency as in~\tabref{table:denoising_snr}.
Even though the model is never trained in SNRs $<$ 0dB, it still performs significantly better than the baseline methods.

\vspace{1pt}
\noindent\textbf{Evaluation results on phase-related objective measurements.}
\tabref{table:phase_metric} summarizes the results using the phase-related metrics, UnRMSE~\cite{gaich2015speech2}, GD and IF~\cite{gaich2015speech1}.
The lower UnRMSE score indicates more similarity to the ground-truth speech with respect to phase variance.
The lower GD and IF scores imply smaller differences between the GD and IF of the enhanced speech than the ground-truth speech.
The results confirm that our method is effective on phase reconstruction.
Although overall enhancement results didn't get higher score compared to the test noisy set on the GD metric, our proposed approach nevertheless achieved better results than the baseline model and the model with only PL.

\input{tables/table_db}
\vspace{-5pt}
\input{tables/table.tex}

%% file: tables/table_snr.tex
\begin{table*}[]
\centering
\caption{Objective measurements of speech enhancement performance in VoiceBank-DEMAND dataset.}\vspace{1pt}
\footnotesize
\begin{tabular}{l || c | C{1.1cm}|C{1.1cm}|C{1.1cm}|C{1.1cm}|C{1.1cm}|C{1.1cm}|c|c|C{1.1cm}}
\toprule
     \bf Methods   & \multicolumn{1}{c|}{\bf WB-PESQ} & \multicolumn{1}{c|}{\bf STOI} & \multicolumn{1}{c|}{\bf ESTOI} & \multicolumn{1}{c|}{\bf SDRi} & \multicolumn{1}{c|}{\bf CSIG} & \multicolumn{1}{c|}{\bf CBAK} & \multicolumn{1}{c|}{\bf COVL} & \multicolumn{1}{c|}{\bf SNRseg} & \multicolumn{1}{c|}{\bf fwSNRseg} & \multicolumn{1}{c}{\bf NCM} \\ 
\midrule\midrule
Noisy   & 1.971  & 0.917 & 0.741 &- & 3.340 & 2.447 & 2.629  & 1.731 & 10.974 & 0.945             \\
\midrule
DEMUCS  & 2.916 & 0.946 & 0.826 & 8.802 & 4.320  & 3.414 & 3.641 & 8.209 & 16.309  & 0.986   \\
+PL     & 2.930 & 0.946 & \bf0.830 & 8.718 & 4.322 &  3.425 & 3.648 & 8.301  &  16.250 & \bf 0.987  \\
+PL+PCL & \bf 2.983  &  \bf 0.947 & 0.827 &\bf 8.989   & \bf 4.343 & \bf 3.449  & \bf 3.685  & \bf 8.358  & \bf 16.328 & \bf 0.987\\
\bottomrule
\end{tabular}
\label{table:denoising}
\vspace{-3pt}
\end{table*}

\begin{table*}[]
\vspace{-15 pt}
\centering
\caption{Speech enhancement results in VoiceBank-DEMAND dataset in various SNRs.}\vspace{1pt}
\footnotesize
\begin{tabular}{l||cccc|cccc|cccc}
\toprule
\multirow{2}{*}{}        & \multicolumn{4}{c|}{\bf WB-PESQ}            & \multicolumn{4}{c|}{\bf STOI}            & \multicolumn{4}{c}{\bf SDRi}            \\
\midrule\midrule
        & 2.5 dB & 7.5 dB & 12.5 dB & 17.5 dB & 2.5 dB & 7.5 dB & 12.5 dB & 17.5 dB & 2.5 dB & 7.5 dB & 12.5 dB & 17.5 dB \\
\midrule
Noisy   & 1.422  & 1.764  & 2.103   & 2.602   & 0.864  & 0.914  & 0.936   & 0.954   & -      & -      & -       & -       \\
\midrule
DEMUCS  & 2.413  & 2.832 & 3.089 & 3.335 &   0.916  &  0.945 & 0.957  & 0.964   & 12.513   & 10.334  & 7.464   &  4.846       \\
+PL     & 2.435 & 2.890 & 3.115 & 3.382 &  0.916 & \bf 0.946 & \bf 0.958 & \bf 0.966 & 12.562 & 10.487 & 7.465 & 4.665 \\
+PL+PCL & \bf 2.469  & \bf 2.915  & \bf 3.142 & \bf 3.412 & \bf 0.918  & \bf 0.946  & \bf 0.958   & 0.965 & \bf 12.730 & \bf 10.650 &  \bf 7.658   & \bf 4.865  \\
\bottomrule
\end{tabular}
\label{table:denoising_snr}
\vspace{-14pt}
\end{table*}

%% file: tables/table_db.tex
\begin{table}[!t]
\vspace{-7pt}
\centering
\caption{Objective measurements of speech enhancement performance on the VoiceBank-DEMAND dataset~(-5dB).}
\footnotesize
\begin{tabular}{l ||C{1.3cm} | C{1.3cm}| C{1.3cm}}
\toprule
     \bf Methods    & \multicolumn{1}{c|}{\bf WB-PESQ} 
                    & \multicolumn{1}{c|}{\bf STOI} 
                    & \multicolumn{1}{c}{\bf SDRi}\\ 
        \midrule\midrule
        Noisy   & 1.200 & 0.802 & -\\
        \midrule
        DEMUCS  & 1.975 & 0.877 & 15.188 \\
        +PL     & 1.978 & 0.879 & 15.291 \\
        +PL+PCL (proposed) & \bf 2.027 &  \bf 0.879 & \bf 15.450 \\
    \bottomrule
\end{tabular}
\label{table:5dB_result}
\vspace{-10pt}
\end{table}

%% file: tables/table.tex
\begin{table}[t]
\vspace{-7pt}
\centering
\caption{Phase-related measurements on the VoiceBank-DEMAND dataset. 
A lower score is better.
}
\vspace{3pt}
\footnotesize
\begin{tabular}{l ||C{1.0cm} | C{1.0cm}|C{1.0cm}}
\toprule
     \bf Methods   & \multicolumn{1}{c|}{\bf UnRMSE} & \multicolumn{1}{c|}{\bf GD} & \multicolumn{1}{c}{\bf IF} \\ 
\midrule\midrule
Noisy   & 6.053 & 0.001 & 0.024     \\
\midrule
DEMUCS  & 4.667 & 0.002 &  0.025 \\
+PL     & 4.769 & 0.002 & 0.023  \\
+PL+PCL (proposed) & \bf 4.633 & \bf 0.001 & \bf 0.022   \\ 
\bottomrule
\end{tabular}
\label{table:phase_metric}
\vspace{-4,5pt}

\end{table}

%% file: 6_conclusion.tex
\section{Conclusion}
\label{sec:conclusion}

In this paper, we introduced a novel phase reconstruction strategy for neural speech enhancement by incorporating phase continuity information into a loss function for model training.
To define our PCL, we built a phase kernel that reflects the derivatives of the phase spectrum across the time and frequency axes, which represent instantaneous frequency and group delay information.
We applied our loss to training a state-of-the-art neural speech enhancement model that originally  uses only L1 reconstruction loss and multi-resolution magnitude spectrum loss.
Experimental results on various noisy data at different SNR levels and phase-related criteria results confirmed that our approach was more effective than the baseline.
Notably, we found that our approach perform better even in low SNR cases and unseen harsh condition, in which phase estimation is known to be difficult.
Our learning strategy can be applied to any type of neural speech enhancement model that utilizes complex spectrograms or waveforms for training.
\break

%% file: 7_acknowledge.tex
\noindent\textbf{Acknowledgements.} 
This research was sponsored by Naver Corporation.
\vspace{-5 pt}

%% file: 0_main.bbl
\begin{thebibliography}{10}

\bibitem{paliwal2003usefulness}
{K. K. Paliwal, and L. Alsteris},
\newblock ``Usefulness of phase in speech processing,''
\newblock in {\em Proc. IPSJ Spoken Language Processing Workshop, Gifu, Japan},
  2003, pp. 1--6.

\bibitem{schluter2001using}
{R. Schluter, and H. Ney},
\newblock ``Using phase spectrum information for improved speech recognition
  performance,''
\newblock in {\em ICASSP}, 2001.

\bibitem{lindgren2003speech}
{A. C. Lindgren, and M. T. Johnson, and R. J. Povinelli},
\newblock ``Speech recognition using reconstructed phase space features,''
\newblock in {\em ICASSP}, 2003.

\bibitem{erdogan2015phase}
{H. Erdogan, and J. R. Hershey, and S. Watanabe, and J. Le Roux},
\newblock ``Phase-sensitive and recognition-boosted speech separation using
  deep recurrent neural networks,''
\newblock in {\em ICASSP}, 2015.

\bibitem{stoller2018wave}
{D. Stoller, and S. Ewert, and S. Dixon},
\newblock ``Wave-u-net: A multi-scale neural network for end-to-end audio
  source separation,''
\newblock in {\em ISMIR}, 2018.

\bibitem{lee2019joint}
{J. Lee, and H. G. Kang},
\newblock ``A joint learning algorithm for complex-valued tf masks in deep
  learning-based single-channel speech enhancement systems,''
\newblock {\em IEEE/ACM Transactions on Audio Speech and Language Processing},
  vol. 27, no. 6, pp. 1098--1108, 2019.

\bibitem{zhang2021weighted}
{J. Zhang, and M. D. Plumbley, and W. Wang},
\newblock ``Weighted magnitude-phase loss for speech dereverberation,''
\newblock in {\em ICASSP}, 2021.

\bibitem{yin2020phasen}
{D. Yin, and C. Luo, and Z. Xiong, and W. Zeng},
\newblock ``Phasen: A phase-and-harmonics-aware speech enhancement network,''
\newblock in {\em AAAI}, 2020.

\bibitem{mowlaee2014interspeech}
{P. Mowlaee, and R. Saeidi, and Y. Stylanou},
\newblock ``Interspeech 2014 special session: Phase importance in speech
  processing applications,''
\newblock in {\em INTERSPEECH}, 2014.

\bibitem{yamamoto2020parallel}
{R.Yamamoto, and E. Song, and J. M. Kim},
\newblock ``Parallel wavegan: A fast waveform generation model based on
  generative adversarial networks with multi-resolution spectrogram,''
\newblock in {\em ICASSP}, 2020.

\bibitem{defossez2020real}
{A. Defossez, and G. Synnaeve and Y. Adi},
\newblock ``Real time speech enhancement in the waveform domain,''
\newblock in {\em INTERSPEECH}, 2020.

\bibitem{alsteris2006further}
{L. D. Alsteris, and K. K. Paliwal},
\newblock ``Further intelligibility results from human listening tests using
  the short-time phase spectrum,''
\newblock {\em Speech Communication}, vol. 48, no. 6, pp. 727--736, 2006.

\bibitem{paliwal2011importance}
{K. Paliwal, and K. W{\'o}jcicki, and B. Shannon},
\newblock ``The importance of phase in speech enhancement,''
\newblock {\em Speech Communication}, vol. 53, no. 4, pp. 465--494, 2011.

\bibitem{williamson2015complex}
{D. S. Williamson, and Y. Wang, and D. Wang},
\newblock ``Complex ratio masking for monaural speech separation,''
\newblock {\em IEEE/ACM Transactions on Audio Speech and Language Processing},
  vol. 24, no. 3, pp. 483--492, 2015.

\bibitem{hu2020dccrn}
{Y. Hu, and Y. Liu, and S. Lv, and M. Xing, and S. Zhang, and Y. Fu, and J. Wu,
  and B. Zhang and L. Xie},
\newblock ``Dccrn: Deep complex convolution recurrent network for phase-aware
  speech enhancement,''
\newblock in {\em INTERSPEECH}, 2020.

\bibitem{venkataramani2017adaptive}
{S. Venkataramani, and J. Casebeer, and P. Smaragdis},
\newblock ``Adaptive front-ends for end-to-end source separation,''
\newblock in {\em NIPS}, 2017.

\bibitem{sun2021funnel}
{Y. Sun, and L. Yang, and H. Zhu, and J. Hao},
\newblock ``Funnel deep complex u-net for phase-aware speech enhancement,''
\newblock in {\em INTERSPEECH}, 2021.

\bibitem{wang2015deep}
{Y. Wang, and D. L. Wang},
\newblock ``A deep neural network for time-domain signal reconstruction,''
\newblock in {\em ICASSP}, 2015.

\bibitem{su2020hifi}
{J. Su, and Z. Jin, and A. Finkelstein},
\newblock ``Hifi-gan: High-fidelity denoising and dereverberation based on
  speech deep features in adversarial networks,''
\newblock in {\em INTERSPEECH}, 2020.

\bibitem{fu2019metricgan}
{S. W. Fu, and C. F. Liao, and Y. Tsao, and S. D. Lin},
\newblock ``Metricgan: Generative adversarial networks based black-box metric
  scores optimization for speech enhancement,''
\newblock in {\em ICML}, 2019.

\bibitem{mowlaee2016advances}
{P. Mowlaee, and R. Saeidi, and Y. Stylianou},
\newblock ``Advances in phase-aware signal processing in speech
  communication,''
\newblock {\em Speech Communication}, vol. 81, pp. 1--29, 2016.

\bibitem{valentini2017noisy}
{C. Valentini-Botinhao, and others},
\newblock ``Noisy speech database for training speech enhancement algorithms
  and tts models,''
\newblock 2017.

\bibitem{hu2007evaluation}
{Y. Hu, and P. C. Loizou},
\newblock ``Evaluation of objective quality measures for speech enhancement,''
\newblock {\em IEEE/ACM Transactions on Audio Speech and Language Processing},
  vol. 16, no. 1, pp. 229--238, 2007.

\bibitem{gaich2015speech1}
{A. Gaich, and P. Mowlaee},
\newblock ``On speech quality estimation of phase-aware single-channel speech
  enhancement,''
\newblock in {\em ICASSP}, 2015.

\bibitem{gaich2015speech2}
{A. Gaich, and P. Mowlaee},
\newblock ``On speech intelligibility estimation of phase-aware single-channel
  speech enhancement,''
\newblock in {\em INTERSPEECH}, 2015.

\end{thebibliography}
